\documentstyle[12pt]{article}

\advance \topmargin by -\headheight
\advance \topmargin by -\headsep

\evensidemargin \oddsidemargin
\begin{document}

\newpage
\pagestyle{empty}

\begin{flushright}
CERN-TH/99-303\\
gr-qc/9910023 \\
October 1999
\end{flushright}

\begin{centering}

\baselineskip 21pt plus 0.2pt minus 0.2pt

\bigskip
{\Large {\bf On the Salecker-Wigner limit and the use of
interferometers in space-time-foam studies}}

\bigskip

\baselineskip 12pt plus 0.2pt minus 0.2pt

\bigskip
\bigskip

{\bf Giovanni AMELINO-CAMELIA}\footnote{{\it Marie Curie
Fellow} (address from February 2000: Dipartimento di Fisica,
Universit\'a di Roma ``La Sapienza'',
Piazzale Moro 2, Roma, Italy)}\\
\bigskip
Theory Division, CERN, CH-1211, Geneva, Switzerland

\end{centering}
\vspace{1.2cm}

\centerline{\bf ABSTRACT }

\begin{quote}

The recent paper gr-qc/9909017
criticizes the limit on the measurability
of distances that was derived by Salecker and Wigner in the 1950s.
If justified, this criticism would have 
important implications for all the
recent studies that have used in various ways the 
celebrated Salecker-Wigner result,
but I show here that
the analysis reported in gr-qc/9909017 is incorrect.
Whereas Salecker and Wigner sought an operative definition
of distances suitable for the Planck regime,
the analysis in gr-qc/9909017 relies on several assumptions
that appear to be natural in the context of
most present-day experiments
but are not even meaningful in the Planck regime.
Moreover, contrary to the claim made in gr-qc/9909017, 
a relevant quantum uncertainty
which is used in the Salecker-Wigner derivation cannot be 
truly eliminated;
unsurprisingly, it can only be traded for another comparable 
contribution to the total uncertainty in the measurement.
I also comment on the role played by the Salecker-Wigner limit in
my recent proposal of interferometry-based tests of quantum
properties of space-time, which was incorrectly described in
gr-qc/9909017.
In particular, I emphasize that,
as discussed in detail in gr-qc/9903080,
only some of the quantum-gravity ideas that can
be probed with modern interferometers are motivated by the 
Salecker-Wigner limit. The bulk of the insight we can expect from
such interferometric studies concerns the properties
of "foamy" models of space-time, which are intrinsically interesting
independently of the Salecker-Wigner limit.

\end{quote}
\baselineskip 18pt

\vfill

\newpage
\pagenumbering{arabic}
\setcounter{page}{1}
\pagestyle{plain}
\baselineskip 12pt plus 0.2pt minus 0.2pt

\section{INTRODUCTION}

In the recent paper~\cite{anos},
Adler, Nemenman, Overduin and 
Santiago criticize a limit on the measurability
of distances which was originally
derived by Salecker and Wigner in the 1950s~\cite{wign}.
If correct, this criticism would have implications for all the recent
papers which have used in one way or another the
celebrated Salecker-Wigner study.
In particular, some of quantum-gravity ideas
that can be tested using the interferometry-based experiments I
proposed in Refs.~\cite{gacgwi,bignapap} are motivated by
the Salecker-Wigner limit; moreover,
the Salecker-Wigner limit is the common ingredient
(even though this ingredient was used in very different ways and
from very different viewpoints~\cite{bignapap,gacmpla})
of several recent studies
concerning possible limitations on the measurability of 
distances~\cite{gacmpla,ng,gacgrf98}
or limitations in the ``tightness'' achievable~\cite{diosi}
in the operative definition of a network of geodesics.

I show here that the analysis reported in Ref.~\cite{anos}
is incorrect. It relies
on assumptions which cannot be justified in the
framework set up by Salecker and Wigner (while the same assumptions
would be reasonable in the context of certain measurements using
rudimentary experimental setups).
In particular, contrary to the claim made in Ref.~\cite{anos}, 
the source of $\sqrt{T_{obs}}$ uncertainty
(with $T_{obs}$ denoting the time of
observation in a sense which will be clarified in the following)
that was considered by Salecker and Wigner cannot be 
truly eliminated;
unsurprisingly, it can only be traded for another
source of $\sqrt{T_{obs}}$ uncertainty.
The analysis reported in Ref.~\cite{anos} also handles
inadequately the idealized concept of ``clock'' relevant for
the type of ``in-principle analysis'' discussed
by Salecker and Wigner.

In addition to this incorrect criticism
of the limit derived by Salecker and Wigner,
Ref.~\cite{anos} also misrepresented the role
of the Salecker-Wigner
limit in providing motivation
for the mentioned
proposal~\cite{gacgwi,bignapap}
of interferometry-based space-time-foam studies.
The reader unfamiliar with the relevant literature
would come out of reading Ref.~\cite{anos}
with the impression that such interferometry-based tests
could only be sensitive to quantum-gravity ideas
motivated by the Salecker-Wigner limit;
instead\footnote{This was already discussed in
detail in Ref.~\cite{bignapap}
which appeared six moths before Ref.~\cite{anos}
but was not mentioned (or taken into account in any other way)
in Ref.~\cite{anos}.}
only some of the quantum-gravity ideas that can
be probed with modern interferometers are motivated by the 
Salecker-Wigner limit. The bulk of the insight we can expect from
such interferometric studies concerns the stochastic properties
of "foamy" models of space-time, which are intrinsically interesting
independently of the Salecker-Wigner limit.

I shall articulate this Letter in sections,
each making one conceptually-independent and simple point. 
I start in the next Section~2 by reviewing which type of ideas
concerning stochastic properties
of "foamy" models of space-time
can be tested with modern interferometers.
From the discussion it will be clear that
interest in these ``foamy" models of space-time
is justified quite independently of
the Salecker-Wigner limit (in fact, this limit
will not even be mentioned in Section~2).
Section~2 is perhaps the most
important part of this Letter,
since its primary objective
is the one of making sure that experimentalists
do not loose interest in the proposed interferometry-based tests
as a result of the confusion generated by Ref.~\cite{anos}.

The remaining sections do concern the Salecker-Wigner limit,
reviewing some relevant results and clarifying various
incorrect statements provided in Ref.~\cite{anos}.
Section~3 briefly reviews the argument
put forward by Salecker and Wigner.
Section~4 emphasizes that the Salecker-Wigner limit is obtained
in ordinary quantum mechanics, but it can provide
motivation for a certain type of ideas concerning
quantum properties of space-time.
The nature of the idealized clock relevant for the type of
analysis performed by Salecker and Wigner
is discussed in Section~5, also clarifying in which sense
some comments on this clock that were made
in Ref.~\cite{anos} are incorrect.
Section~6 clarifies how the potential well
considered in Ref.~\cite{anos} would simply trade one source
of $\sqrt{T_{obs}}$ uncertainty
for another source of $\sqrt{T_{obs}}$ uncertainty.
In Section~7 I clarify that the comments on
decoherence of the clock presented in Ref.~\cite{anos}
would not apply to the Salecker-Wigner setup.
Section~8 is devoted to some closing remarks.

\section{FOAMY \space SPACE-TIME \space AND  \space MODERN
 \space $~$  \space $~$  \space $~$  \space
 INTERFEROMETERS}

A prediction of nearly
all approaches to the unification
of gravitation and quantum mechanics is that 
at very short distances the sharp
classical concept of space-time should give way 
to a somewhat ``fuzzy'' (or ``foamy'') picture,
possibly involving virulent
geometry fluctuations (sometimes intuitively/heuristically
associated with virtual black holes and wormholes).
This subject originates from observations
made by Wheeler~\cite{wheely}
and Hawking~\cite{hawk}
and has developed into a rather vast literature.
Examples of recent proposals in this area
(and good starting points for a literature search)
can be found in
Refs.~\cite{arsarea,peri,fotinilee,gacgrb,gampul,fordlightcone},
which explored possible implementations/consequences of space-time
foam ideas in various versions of quantum gravity, and in
Refs.\cite{emn,aemn1,adrian}, which performed similar studies
in an approach based on non-critical strings.
Although the idea of space-time foam appears to have significantly
different incarnations in different quantum-gravity approaches,
a general expectation that emerges from this framework is that 
the distance between two bodies ``immerged" in the space-time foam
would be affected by (quantum-gravity-induced) fluctuations.

A phenomenological model of fluctuations affecting
a quantum-gravity distance must describe
the underlying stochastic processes.
As explained in detail in Refs.~\cite{bignapap,polonpap},
from the point of view of comparison with data 
obtainable with modern interferometers
the best way to characterize such models is through
the associated amplitude spectral density of
distance fluctuations~\cite{amplspectdef,saulson}.
A natural starting point for the parametrization
of this amplitude spectral density 
is given by\footnote{Of course, 
a parametrization such as the one in
Eq.~(\ref{gacspectrbeta})
could only be valid for frequencies $f$ significantly
smaller than the Planck frequency $c/L_{p}$
and significantly larger than the inverse of the time scale
over which the classical geometry 
of the space-time region where
the experiment is performed manifests
significant curvature effects.}
\begin{eqnarray}
S(f)=f^{-\beta} \, ({\cal L}_{\beta})^{{3 \over 2}-\beta}
\, c^{\beta-{1 \over 2}} ~,
\label{gacspectrbeta}
\end{eqnarray}
where $c$ is the speed-of-light constant, the
dimensionless parameter $\beta$
carries the information on the nature of the underlying stochastic
processes and the dimensionful
(length) parameter ${\cal L}_{\beta}$ carries
the information on the magnitude and rate of the
fluctuations\footnote{I am assigning an
index $\beta$ to ${\cal L}_{\beta}$ just in order to facilitate
a concise description of experimental bounds.
For example,
if data were to rule out 
the fluctuations scenario with, say, $\beta = 0.6$
for all values of the effective length 
scale down to, say, $10^{-27}m$
one could simply write the
formula ${\cal L}_{\beta= 0.6} < 10^{-27}m$.}.
A detailed discussion of the definition and applications
of this type of amplitude spectral density
can be found in Ref.~\cite{amplspectdef,saulson}.
For the readers unfamiliar with the use of amplitude spectral
densities some useful intuition can be obtained
from the fact that~\cite{amplspectdef,rwold}
the standard deviation of the fluctuations is
formally related to $S(f)$ by
\begin{eqnarray}
\sigma^2 = \int_{1/T_{obs}}^{f_{max}}
[S(f)]^2 \,  df ~,
\label{gacspectrule}
\end{eqnarray}
where $T_{obs}$ is the time over which the distance is kept
under observation.

In Eq.~(\ref{gacspectrbeta})
the parameter $\beta$ could in principle take any value,
and it is even quite plausible that 
in reality the stochastic processes
(if at all present) would have a more
complex structure than the simple power law
codified in Eq.~(\ref{gacspectrbeta}). Still,
Eq.~(\ref{gacspectrbeta}) appears to be the natural starting
point for a phenomenological programme of exploration
of the possibility of ``distance-fuzziness'' effects induced
by quantum properties of space-time.
In particular, it seems natural to devote special attention
to values of $\beta$ in the range $1/2 \le \beta \le 1$;
in fact, as explained in greater detail in Refs.~\cite{bignapap},
$\beta = 1/2$ is the type of behaviour one would expect~\cite{jare}
in fuzzy space-times without quantum decoherence
(without ``information loss''),
while the case $\beta = 1$
provides the simplest model
of stochastic (quantum) fluctuations
of distances,
in which a distance is affected by completely random minute
(possibly Planck-length size) fluctuations which
can be modeled as stochastic processes of random-walk type.
Values of $\beta$ somewhere in between the 
cases $\beta = 1/2$ and $\beta = 1$ could provide a
rough model of space-times with decoherence
effects somewhat milder than the $\beta = 1$ random-walk case.
In other words, in light of the realization~\cite{bignapap,jare}
that foamy space-times without decoherence would only be consistent
with distance fluctuations of type
$\beta = 1/2$ the popular arguments that support
quantum-gravity-induced deviations from quantum coherence
motivate interest in values of $\beta$ somewhat different from $1/2$.

Readers unfamiliar with the subject can get an intuitive picture
of the relation between the value of $\beta$ and decoherence
by resorting again to Eq.~(\ref{gacspectrule}).
For example, as discussed in greater detail in
Ref.~\cite{bignapap,rwold},
the case $\beta = 1$ corresponds to $\sigma \sim \sqrt{T_{obs}}$,
the standard deviation characteristic of random-walk processes,
and this type of $T_{obs}$-dependence would be consistent
with decoherence in the sense that the information stored
in a network of distances would degrade over
time\footnote{For example,
an observer could store ``information''
in a network of bodies by adjusting their
distances to given values at a given initial time.
If space-time did involve distance fluctuations with standard
deviation that grows with the time of observation,
there would be an intrinsic mechanism for
this information to degrade over time.
Other intuitive descriptions of the relation between
certain fuzzy space-times and decoherence
can be found in Ref.~\cite{ng}.
Depending on the reader's background
it might also be useful to adopt
the language of the ``memory effect'',
as done, for example, in Ref.~\cite{memory}.}.
Similar observations, but with weaker power-law dependence
on $T_{obs}$, hold for values of $\beta$
in the range $1/2 < \beta < 1$.
In the limiting case $\beta = 1/2$ the $T_{obs}$-dependence
turns from power-law to logarythmic,
and this is of course the closest one can get to modeling
space-times without intrinsic decoherence
({\it i.e.} such that the associated standard deviation
is $T_{obs}$-independent) within the
parametrization set up
in Eq.~(\ref{gacspectrbeta})\footnote{As explained in
Refs.~\cite{gacgwi,bignapap} and reviewed here below,
we are still very far from being able to test the type
fuzziness one might expect for space-times without decoherence.
It is therefore at present quite sufficient to model
this type of fuzziness by taking $\beta = 1/2$
in (\ref{gacspectrbeta}).
Readers with an academic interest in seeing a more complete
description of stochastic processes
plausible for a space-time without decoherence
can consult Ref.~\cite{jare}.}.

As observed in Ref.~\cite{bignapap}, independent support
for a fuzzy picture of space-time of the type here being considered
comes from recent
studies~\cite{aemn1,gacgrb,gampul,fordlightcone,adrian}
suggesting that space-time foam
might induced a deformation of the dispersion relation that
characterizes the propagation of the massless particles
used as space-time probes in the operative definition
of distances.
Such a deformation of the dispersion relation would
affect~\cite{aemn1,gacgrb,bignapap}
the measurability of distances just in the way
expected for a fuzzy picture of space-time
of the type here being considered.

In general the connection between loss of quantum coherence
and a foamy/fuzzy picture of space-time is very deep and
has been discussed in numerous publications
(a sample of recent ideas in this area can be found
in Refs.~\cite{ng,peri,elmn,hpcpt,hawkdeconew}).
However, while a substantial amount
of work has been devoted to the ``physics case''
for quantum-gravity induced decoherence,
enormous difficulties have
been encountered in developing a satisfactory
formalism for this type of quantum gravity.
The primary obstruction for the search of the
correct decoherence-encoding formalism
is the fact that a new mechanics would be needed
(ordinary quantum mechanics evolves pure states into pure states)
and the identification of such a new mechanics
in the absence of any guidance from experiments
is extremely hard.
It is in this context that a phenomenology
based on the parametrization (\ref{gacspectrbeta})
finds its motivation.
When a satisfactory workable formalism implementing the
intuition on quantum-gravity-induced
decoherence becomes available, we will be
in a position to extract from it
a specific form of the stochastic processes characterizing
the associated foamy space-time,
with a definite prediction for $S(f)$.
While waiting for these developments on the theoretical-physics side
we might get some help from experiments; in fact,
as observed in Refs.~\cite{gacgwi,bignapap},
the remarkable sensitivity of modern interferometers
(the ones whose primary objective is the detection
of the classical-gravity phenomenon of gravity
waves~\cite{saulson})
allows us to put significant bounds on the parameters
of Eq.~(\ref{gacspectrbeta}).
While it is remarkable that some candidate
quantum-gravity phenomena
are within reach of doable experiments,
it is instead quite obvious that interferometers
would be the natural in-principle
tools for the study of distance fluctuations.
In fact, the operation of interferometers 
is based on the detection of
minute changes in the positions of some test masses
(relative to the position of a beam splitter),
and, if these positions were affected by
quantum fluctuations of the type discussed
above, the operation of interferometers
would effectively involve an additional
source of noise due to quantum gravity~\cite{gacgwi,bignapap}.

The data obtained at
the {\it Caltech 40-meter interferometer}, which
in particular achieved~\cite{ligoprototype}
displacement noise levels with amplitude spectral density
of about $3 \cdot 10^{-19} m/\sqrt{H\!z}$
in the neighborhood of $450$ $H\!z$, 
allow us to set the bound~\cite{gacgwi,bignapap,polonpap}
\begin{eqnarray}
[{\cal L}_{\beta}]_{Caltech}
< \left[ {3 \cdot 10^{-19} m \over \sqrt{H\!z}}
 \, (450 H\!z)^\beta \,
c^{(1-2\beta)/2} \right]^{2/(3 - 2 \beta)} ~.
\label{boundcalty}
\end{eqnarray}

In order to get some intuition
for the significance of this bound
let us consider the case $\beta = 1$.
For $\beta =1$
the bound in Eq.~(\ref{boundcalty})
takes the form $L_{\beta = 1} < 10^{-40}m$.
This is quite impressive
since $\beta = 1$, $L_{\beta=1} \sim 10^{-35}m$
corresponds to fluctuations in the 40-meter arms of
the Caltech interferometer
that are of Planck-length magnitude ($L_p \sim 10^{-35}m$)
and occur at a rate of one per each
Planck-time interval ($t_p = L_p/c \sim 10^{-44} s$).
The data obtained at
the {\it Caltech 40-meter interferometer}
therefore rule out this simple model in spite of the
minuteness (Planck-length!!)
of the fluctuations involved.
Another intuition-building
observation concerning the significance of this result
is obtained by considering the standard
deviation $\sigma \sim \sqrt{L_p c T_{obs}}$ which
would correspond to such Planck-length
fluctuations occurring at $1/t_p$
rate. From $\sigma \sim \sqrt{L_p c T_{obs}}$
one predicts fluctuations with standard deviation
even smaller than $10^{-5}m$ on a time of observation as large
as $10^{10}$ years (the size of the whole observable universe
is about $10^{10}$ light years!!) but
in spite of their minuteness these can
be ruled out exploiting the remarkable sensitivity of modern
interferometers.

Additional comments on values of $\beta$
in the range
$1/2 < \beta < 1$ can be found in Refs.~\cite{bignapap,polonpap}
(in Ref.~\cite{bignapap} the reader will find a detailed
discussion of the case $\beta = 5/6$).
In the present Letter it suffices to observe that
the bound encoded in Eq.~(\ref{boundcalty})
becomes less stringent as the value of $\beta$
decreases. In particular, in the limit $\beta = 1/2$,
the case providing an effective model for space-times
without intrinsic decoherence,
Eq.~(\ref{boundcalty}) only implies
${\cal L}_{\beta = 1/2} < 10^{-17}m$,
which is still very comfortably consistent with the
natural expectation~\cite{jare} that within that framework
one would have ${\cal L}_{\beta = 1/2} \sim  L_p \sim 10^{-35}m$.

In this section I have in no way considered
the statements on the Salecker-Wigner limit
reported in Ref.~\cite{anos}.
As anticipated in the Introduction, I have opened the paper
with this section briefly summarizing
the status of interferometry-based studies of
distance fuzziness. The fact that the Salecker-Wigner limit
was not even
mentioned in this section should however clarify that, contrary to
the impression one gets from reading Ref.~\cite{anos},
these interferometric studies are intrinsically interesting,
quite independently of any consideration concerning
the Salecker-Wigner limit.
This is already clear at least to a portion of the community;
for example, in recent work~\cite{adrian} on foamy space-times
(without any reference to the Salecker-Wigner related literature)
the type of modern-interferometer sensitivity exposed in
Refs.~\cite{gacgwi,bignapap} was used
to constrain certain novel candidate light-cone-broadening effects .

The brief review provided in this section should also clarify
in which sense another statement provided in Ref.~\cite{anos}
is misleading. It was in fact stated in Ref.~\cite{anos}
that, since the sensitivity of modern interferometers is at the
level\footnote{For example,
planned interferometers~\cite{ligo,virgo}
with arm lengths of a few $Km$
expect to detect gravity waves of amplitude as 
low as $3 \cdot 10^{-22}$ (at frequencies of about $100 Hz$).
This roughly means that these modern gravity-wave interferometers
should monitor the (relative) positions of their test masses
(the beam splitter and the mirrors)
with an accuracy of order $10^{-18} m$.}
of $10^{-18}m$,
any quantum-gravity model tested by such interferometers
should predict a break down of the classical space-time
picture on distance scales of order $10^{-18}m$.
Let me illustrate in which sense this statement misses the
substance of the proposed tests
by taking again as an example the one 
with $\beta = 1$, which allows an intuitive discussion in terms of
simple random-walk processes.
We have seen that this can describe fluctuations of
Planck-length magnitude occurring at $1/t_p$ rate.
All the scales involved in the stochastic picture are at
the $10^{-35}m$ scale, but we can rule out
this scenario using a ``$10^{-18}m$ machine''
because this machine operates at frequencies
of order a few hundred $Hz$ (which correspond to
time scales of order a few milliseconds)
and therefore is effectively sensitive to the collective effect
of a very large number of minute Planck-scale effects
({\it e.g.}, in the simple random-walk case,
during a time of a few milliseconds as many as $10^{41}$
Planck-length fluctuations would affect the arms of the
interferometer).
This is not different from other similar experiments
probing fundamental physics.
For example, proton-decay
experiments use protons at rest (objects of size $10^{-16}m$)
to probe physics on distance scales of order $10^{-32}m$
(the conjectured size of gauge bosons mediating proton decay),
and this is done by monitoring a very large number of protons
so that the apparatus is sensitive to a collective effect
which is much larger than the decay probability of each
individual proton.
A similar idea has already been exploited in ``quantum-gravity
phenomenology''~\cite{polonpap}; in fact, the experiment
proposed in Ref.~\cite{gacgrb} is possible only because
the photons that reach us from distant astrophysical sources
have traveled for such a long
time that they are in principle
sensitive to the collective effect
of a very large number of interactions with the
space-time foam.

\section{THE SALECKER-WIGNER LIMIT IN ORDINARY QUANTUM MECHANICS}

Having clarified what part of the motivation for interferometric
studies is completely independent of the Salecker-Wigner limit
I have two remaining tasks: the one of providing
a brief review of
the Salecker-Wigner limit
and the one of correcting the incorrect statements on
the Salecker-Wigner limit which were given in Ref.~\cite{anos}.
Let me start by considering the original Salecker-Wigner limit
within ordinary quantum mechanics.
The analysis reported by Salecker and Wigner in Ref.~\cite{wign}
concerns the measurability of distances.
In particular, they considered
the measurement of the distances defined by the
network of free-falling
bodies that might compose an idealized ``material reference
system''~\cite{rovellimrs}. Those who have been developing
the research line started by Salecker and Wigner
have also considered more general distance measurements,
but the emphasis has remained on measurement analyses
that might provide intuition on the way
in which distances could be in principle operatively defined
in quantum gravity.
The essence
of the Salecker-Wigner argument can be summarized
as follows. They ``measured'' (in the ``{\it gedanken}'' sense)
the distance $D$ between two bodies 
by exchanging a light signal between them.
The measurement procedure requires {\it attaching}\footnote{Of
course, for consistency with causality,
in such contexts one assumes devices to be ``attached non-rigidly,''
and, in particular, the relative position
and velocity of their centers of mass continue to satisfy the
standard uncertainty relations of quantum mechanics.} 
a light-gun ({\it i.e.} a device 
capable of sending
a light signal when triggered), a detector
and a clock to
one of the two bodies 
and {\it attaching} a mirror to the other body. 
By measuring the time $T_{obs}$ (time of observation)
needed by the light signal
for a two-way journey between the bodies one 
also obtains a measurement of  
the distance $D$.
For example, in flat space 
and neglecting quantum effects 
one simply finds that $D = c {T_{obs} / 2}$.
Unlike most conventional measurement analyses,
Salecker and Wigner were concerned with the quantum
properties of the devices involved in the measurement
procedure. In particular, since they were considering
a distance measurement, it was clear that
quantum uncertainties in the position (relative to, say,
the center of mass of the two
bodies whose distance is being measured) of some of the
devices involved in the measurement procedure would translate
into uncertainties in the overall measurement of $D$.
Importantly, the analysis of these device-induced
uncertainties leads to
a lower bound on the measurability of $D$.
To see this it is sufficient to consider the
contribution to $\delta D$ coming from
only one of the quantum uncertainties that affect
the motion of the devices.
In Ref.~\cite{wign} (and in the more recent studies
reported in Refs.~\cite{ng,diosi})
the analysis focused on the uncertainty in the position
of the Salecker-Wigner clock, while in some of my related
studies~\cite{gacmpla,gacgrf98} the analysis focused on 
the uncertainties that affect the motion
of the center of mass of the system
composed by the light-gun, the detector and the clock.
These approaches are actually identical, since (as I shall
discuss in greater detail later) the Salecker-Wigner clock
is conceptualized~\cite{wign} as a device not only capable of
keeping track of time but also
capable of sending and receiving signals; it is therefore
a composite device including at least
a clock, a transmitter and a receiver.
Moreover, the substance of the argument
does not depend very sensitively on which position
is considered, as long as it is associated with a
device whose position must be known over the whole
time required by the measurement procedure.
For definiteness, let me here proceed
denoting with $x^*$ and $v^*$
the position and the velocity of an idealized Salecker-Wigner
clock. Assuming that the experimentalists prepare this device
in a state characterised by
uncertainties $\delta x^*$ and $\delta v^*$,
one easily finds~\cite{wign,gacmpla,ng,gacgrf98}
\begin{eqnarray}
\delta D \geq 
\delta x^* + T_{obs} \delta v^* 
\geq 
\delta x^* 
+ \left( {1 \over  M_b} + {1 \over  M_d} \right)
{ \hbar T_{obs} \over 2 \, \delta x^* }
~,
\label{deltawignOLDprologo}
\end{eqnarray}
where $M_b$ is the sum of the masses of the two bodies
whose distance is being measured, $M_d$ is
the mass of the device 
being considered ({\it e.g.}, the mass of the clock)
and
I also used the fact 
that Heisenberg's {\it Uncertainty Principle} 
implies $\delta x^* \delta v^* \ge (1/M_b + 1/M_d) \hbar/2$.
[The {\it reduced mass} $(1/M_b+1/M_d)^{-1}$ 
is relevant for the relative motion of the clock 
with respect to the position of
the center of mass of the system composed by the two
bodies whose distance is being measured.]

Evidently, from (\ref{deltawignOLDprologo})
it follows that for given $M_b$ and $M_d$ there is a lower bound
on the measurability of $D$
\begin{eqnarray}
\delta D \geq \sqrt{ {\hbar T_{obs} \over 2}
\left( {1 \over  M_b} + {1 \over  M_d} \right) }
~.
\label{deltawignOLD}
\end{eqnarray}

The result (\ref{deltawignOLD})
may at first appear somewhat puzzling, since
ordinary quantum mechanics should not
limit the measurability of any given observable.
[It only limits the combined measurability
of pairs of conjugate observables.]
However, from a physical/phenomenological and conceptual
viewpoint it is well understood that the
proper framework for the application of the formalism
of quantum mechanics is the
description of the results of measurements performed
by classical devices (devices that can be treated
as approximately classical within the level of
accuracy required by the measurement).
It is therefore not surprising
that the infinite-mass (classical-device\footnote{A rigorous 
definition of a ``classical device'' is 
beyond the scope of this Letter. However, it should be emphasized 
that the experimental setups being here considered require
the devices to be accurately positioned during the time
needed for the measurement, and therefore an ideal/classical
device should be infinitely massive so 
that the experimentalists can prepare it in a state 
with $\delta x \, \delta v \sim \hbar/M \sim 0$.})
limit turns
out to be required 
in order to bridge the gap between (\ref{deltawignOLD}) 
and the prediction $min \delta D = 0$
of the formalism
of ordinary quantum mechanics.\footnote{Perhaps
more troubling is the fact that $min \delta D = 0$
appears to require not only an infinitely large $M_d$
but also an infinitely large $M_b$.
One feels somewhat uncomfortable
treating the mass of the bodies whose distance is being
measured as a parameter of the apparatus. This might be
another pointer to the fact that quantum measurement
of gravitational/geometric observables requires
a novel conceptualization of quantum mechanics. I postpone
the consideration of this point to future work.}

In this section on the Salecker-Wigner limit
I have not taken into account the
gravitational properties of the devices.
It has been strictly confined within 
ordinary (non-gravitational) quantum mechanics.
Actually, one can interpret the Salecker-Wigner limit
as one way to render manifest the true nature of the
physical applications of the quantum-mechanics formalism
and its relation with a certain class of experiments
(the ones performed by classical devices).
The picture emerging from the analysis
of Salecker and Wigner fits well
within a general picture
emerging from other similar studies.
In particular,
the celebrated Bohr-Rosenfeld analysis~\cite{rose}
of the measurability of the electromagnetic field
found that the accuracy allowed by the formalism
of ordinary quantum mechanics could only be achieved
using a very special type of device:~idealized
test particles with vanishing ratio between
electric charge and inertial mass.

\section{FROM \space THE \space SALECKER-WIGNER  \space
LIMIT \space TO \space QUANTUM GRAVITY}

Let me now take the Salecker-Wigner limit as starting point
for a quantum-gravity argument. I will therefore now not only
consider the quantum properties of the devices, but also their
gravitational properties.
It is well understood (see, {\it e.g.},
Refs.~\cite{gacmpla,gacgrf98,diosi,bergstac,dharam94grf,dharam3QG}) 
that the combination of  the gravitational properties 
and the quantum properties of devices can have an important role
in the analysis of the operative definition of gravitational
observables.
Actually, by ignoring the way in which the gravitational properties 
and the quantum properties of devices combine in measurements
of geometry-related physical properties of a system
one misses some of the fundamental elements
of novelty we should expect for the interplay of gravitation
and quantum mechanics; in fact, one would be missing an
element of novelty which is deeply associated with the Equivalence
Principle.
For example, 
attempts to generalize the mentioned Bohr-Rosenfeld analysis
to the study of gravitational fields
(see, {\it e.g.}, Ref.~\cite{bergstac})
are of course confronted with the fact that the ratio between
gravitational ``charge'' (mass) and inertial mass
is fixed by the Equivalence Principle.
While ideal devices with vanishing ratio between  
electric charge and inertial mass can
be considered at least in principle,
devices with vanishing ratio between  
gravitational mass and inertial mass 
are not admissible in any (however formal) limit
of the laws of gravitation.
This observation provides one of the strongest elements
in support of the idea~\cite{gacgrf98}
that the mechanics on which quantum
gravity is based must not be exactly
the one of ordinary quantum mechanics.
In turn this contributes to the whole spectrum of arguments
that support the expectation that the loss of quantum coherence
might be intrinsic in quantum gravity.

Similar support for quantum-gravity-induced
decoherence emerges from taking into account both
gravitational and quantum properties of devices in
the analysis of the Salecker-Wigner measurement procedure.
The conflict with ordinary quantum mechanics
immediately arises because the
infinite-mass limit
is in principle inadmissible for measurements
concerning gravitational effects.
As the devices get more and more massive they increasingly 
disturb the gravitational/geometrical observables, and
well before reaching the infinite-mass limit the procedures 
for the measurement of gravitational observables cannot
be meaningfully performed~\cite{gacmpla,ng,gacgrf98}.
These observations, which render unaccessible
the limit of vanishingly small right-hand-side
of Eq.~(\ref{deltawignOLD}),
provide motivation for the possibility~\cite{gacmpla,gacgrf98}
that in quantum gravity
there be a $T_{obs}$-dependent intrinsic uncertainty
in any measurement that monitors
a distance $D$ for a time $T_{obs}$.
Gravitation forces us to renounce to
the idealization of infinitely-massive devices
and this in turn forces us to deal with the element of
decoherence encoded in the fact that measurements
requiring longer times of observation are
intrinsically/fundamentally
affected by larger quantum uncertainty.

It is important to realize that this
element of decoherence found in the analysis of
the measurability of distances
comes simply from combining elements of quantum mechanics
with elements of classical gravity.
As it stands it is not to be interpreted as a
genuine quantum-gravity effect, but of course
this argument based on the Salecker-Wigner limit
provides motivation for the exploration of the possibility
that quantum gravity might accommodate this type of
decoherence mechanism at the fundamental level.
In the analysis of the Salecker-Wigner setup
the $T_{obs}$ dependence is not introduced at the fundamental
level; it is a derived property emerging from
the postulates of gravitation and quantum mechanics.
However, it is plausible that quantum gravity,
as a fundamental theory of space-time, might
accommodate this type of bound at the fundamental
level ({\it e.g.}, among its postulates or as a straightforward
consequence of the correct short-distance picture of space-time).
It is through this
(plausible, but, of course, not self-evident)
argument that the Salecker-Wigner
limit provides additional motivation for the interferometric studies
discussed in Section~2.
The element of decoherence encoded in the stochastic models of
fuzzy space-time is quite consistent with the type of decoherence
mechanism suggested by the analysis of the Salecker-Wigner 
measurement procedure.
One could see the Wheeler-Hawking picture of an ``active''
quantum-gravity vacuum and the measurability bound
suggested by the analysis of the Salecker-Wigner 
measurement procedure
as independent arguments in support of distance fuzziness
of the type here reviewed in Section~2.
Of course, the intuition associated to the arguments
of Wheeler, Hawking and followers is more fundamental
and has wider significance, but the 
analysis of the Salecker-Wigner 
measurement procedure has the advantage of allowing
to develop (however heuristic) arguments in support
of one or another form of fuzziness, whereas the lack of
explicit models providing a satisfactory implementation of the
Wheeler-Hawking intuition forces one to adopt parametrizations
as general as the one in Eq.~(\ref{gacspectrbeta}).
From this point of view, arguments based on the
Salecker-Wigner measurement procedure can play a role similar
to the one played by the arguments based on
quantum-gravity-induced deformations of dispersion relations,
which, as already mentioned in Section~2,
can also be used~\cite{bignapap} to support specific
corresponding models of fuzziness (values of $\beta$)
within the class of
models parametrized in Eq.~(\ref{gacspectrbeta}).

Let me devote the rest of this section to some of the
arguments based on analyses of the Salecker-Wigner 
measurement procedure that provide support
for one or another form of distance fuzziness.
As observed in Refs.~\cite{gacgwi,bignapap}
a particular value of $\beta$ can be motivated by
arguing in favour of a corresponding
explicit form of the $T_{obs}$ dependence
of the bound on the measurability of distances.
Let me here emphasize that
the robust part of the quantum-gravity argument
based on the analysis of the Salecker-Wigner 
measurement procedure only allows one to conclude
that the $T_{obs}$ dependence cannot be eliminated,
and this is not sufficient for obtaining an explicit
prediction for the $T_{obs}$-dependent measurability
bound. A robust derivation of such an explicit formula
would require one to have available the correct quantum
gravity and derive from it whatever quantity turns out
to play effectively the role of the minimum
quantum-gravity-allowed value of $M_b^{-1}+M_d^{-1}$.
Since quantum gravity is not available to us,
we can only attempt intuitive/heuristic
answers to questions such as:
should quantum gravity host such an effective
minimum value of $M_b^{-1}+M_d^{-1}$?
how small could this effective
minimum value of $M_b^{-1}+M_d^{-1}$ be?
could this minimum value depend on $T_{obs}$?
could it depend on the distance scales being probed?
This questions are discussed in detail in
Refs.~\cite{bignapap,gacmpla,gacgrf98,polonpap}.
For the objectives of the present Letter
it is important to discuss explicitly in which sense
one is seeking answers to these questions.
In seeking these answers one is trying to
get an intuition for the fundamental
conceptual structure of quantum gravity, and therefore
one considers the measurement
procedure from a viewpoint that would
seem appropriate for the definition of distances
possibly as short as the Planck length.
[Some authors (quite reasonably)
would also expect quantum gravity
to accommodate some sort of operative definition of
space-time based on a network of material-particle
(possibly minute clocks) worldlines.]
It is from these viewpoints that one must approach
the questions raised by analyses
of the Salecker-Wigner setup.
As it will be discussed in the next three sections,
one is led to very naive conclusions by adopting instead
a conventional viewpoint based on the intuition
that comes from present-day rudimentary (from a Planck-length
perspective) experimental setups.
The logic of the line of research started by the work
of Salecker and Wigner is the one of applying
the language/structures we ordinarily use in physical contexts
we do understand to contexts that instead seem to lie in
the realm of quantum gravity, hoping that this might guide us
toward some features of the correct quantum gravity.
We already know the answers to the above questions within
ordinary gravitation and quantum mechanics, and therefore an
exercise such as the one reported in Ref.~\cite{anos}
could not possibly teach us anything. It is instead at least
plausible that we get a glimpse of a true property of quantum
gravity by exploring the consequences of removing one of
the elements of the ordinary conceptual structure of quantum
mechanics. The Salecker-Wigner study
(just like the Bohr-Rosenfeld analysis)
suggests that among these conceptual elements of
quantum mechanics the one that is most likely
(although there are of course no guarantees)
to succumb to the unification of gravitation and quantum
mechanics is the requirement for devices to be treated
as classical. Removal of this requirement appears to guide
us toward some candidate properties of quantum gravity (not
of the ordinary laws of gravitation and quantum mechanics!),
which we can then hope to test directly in the laboratory
(as in some cases is actually possible~\cite{gacgwi,bignapap}).

I shall go back to these important points in the next three
sections, but before I do that let me just
briefly summarize the outcome of two simple
attempts to extract quantum-gravity intuition from
the conceptual framework set up by Salecker and Wigner.
One of these approaches I have developed in
Refs.~\cite{bignapap,gacmpla,gacgrf98}.
It is based on the simple observation
that if in quantum gravity the effective
minimum value of $M_b^{-1}+M_d^{-1}$
was $T_{obs}$-independent and $\delta D$-independent,
say $min (M_b^{-1}+M_d^{-1})
= [max(M^*)]^{-1} \equiv c L_{QG}/\hbar$,
we would then get
a bound on the measurability of distances
which goes like $\sqrt{T_{obs}}$
\begin{eqnarray}
\delta D \geq \sqrt{ {\hbar T_{obs} \over 2 \, max(M^*)}}
\equiv \sqrt{ {c T_{obs} L_{QG} \over 2}}
~,
\label{deltawignGACm}
\end{eqnarray}
and would therefore be
suggestive~\cite{gacgwi,bignapap,polonpap}
of random-walk stochastic processes.
I also observed that, if this effective $max(M^*)$
of quantum gravity could still be interpreted as some
maximum mass of the devices used in the measurement
procedure, the value of $max(M^*)$ could be bound by the
observation that in order to allow the measurement procedure
to be performed these devices should at least be light
enough not to turn into black holes.
This allows one to trade~\cite{bignapap,gacmpla,gacgrf98}
the effective mass scale $max(M^*)$
for an effective length scale $s^*$
which would be the maximum effective size\footnote{From
the viewpoint clarified above it is natural to envision
that this length scale $s^*$ would be a fundamental
scale of quantum gravity. Instead of introducing a dedicated scale
for it one could be tempted to consider the possibility that $s^*$
be identified with the only known quantum-gravity scale $L_p$,
even though this would render somewhat daring
the possible interpretation of $s^*$
as maximum size of the devices involved in the measurement.
In a sense more precisely discussed
in Refs.~\cite{gacgwi,bignapap,polonpap}, this identification
$s^* \equiv L_p$ is already ruled out by the same Caltech data
mentioned above~\cite{ligoprototype}.}
allowed in quantum gravity for the individual devices 
partecipating to the measurement procedure:
\begin{eqnarray}
\delta D \geq \sqrt{ {L_p^2 \, c \, T_{obs} \over s^*}}
~.
\label{deltawignGACs}
\end{eqnarray}
[Of course, this whole exercise of trading $max(M^*)$
for $s^*$ only serves the purpose of giving an
alternative intuition for the new length scale $L_{QG}$,
which can now be seen as related to some effective maximum size
of devices $s^*$ by the equation $L_{QG} \equiv L_p^2/s^*$.]

Another approach to the derivation of a candidate quantum-gravity
bound on the measurability of distances 
from an analysis of the Salecker-Wigner 
measurement procedure has been developed by Ng
and Van Dam~\cite{ng}.
These authors took a somewhat different definition
of measurability bound~\cite{bignapap,gacmpla,gacgrf98}
and they also advocated a certain classical-gravity
approach to the estimate of $max(M^*)$.
The end result was 
\begin{eqnarray}
\delta D \geq (L_p^2 \, c \, T_{obs})^{1/3}
~.
\label{deltawignNG}
\end{eqnarray}
In Ref.~\cite{gacgwi,bignapap}
it was observed that a $T_{obs}$-dependence of
the type in Eq.~(\ref{deltawignNG}) would be suggestive
of the stochastic space-time model
with $\beta = 5/6$.

It is interesting to observe~\cite{ng,gacmpla}
that relations such as (\ref{deltawignGACs})
and (\ref{deltawignNG}) can take the form of $D$-dependent
bounds on the measurability of $D$ by observing
that $D \sim T_{obs}$ in typical measurement setups.
The bounds would
be $\delta D \geq \sqrt{D L_{QG}} \equiv \sqrt{D L_p^2/s^*}$
and $\delta D \geq (D L_p^2)^{1/3}$
respectively for (\ref{deltawignGACs})
and (\ref{deltawignNG}).

\section{ON THE SALECKER-WIGNER CLOCK}

As manifest in the brief review provided in the previous two sections,
the Salecker-Wigner limit and the associated intuition concerning
quantum properties of space-time is based on an in-principle analysis
of the measurement of distances, with emphasis on the nature of the
devices used in the measurement procedure.
Accordingly, the measurement procedure is only schematically
described and only from a conceptual point of view.
The devices used in the measurement procedure are also only
considered from the point of
view of the role that they play in the conceptual
structure of the measurement procedure.
For example (an example which is relevant for some of the incorrect
conclusions drawn in Ref.~\cite{anos}), the
Salecker-Wigner ``clock'' is not simply a
timing device, but it is to be intended as the network of
instruments needed for the ``clock'' to play its role
in the measurement
procedure ({\it e.g.} instruments needed to trigger the
transfer of information from the clock to the rest of the network of
devices that form the apparatus or instruments needed to affect the
position of the clock in ways needed by the measurement procedure).
This was already very clearly explained in the early works~\cite{wign}
by Salecker and Wigner, which in various points state
that the relevant idealized clocks are, for example,
capable of sending and receiving signals
(they are therefore composite devices including at least
a clock, a transmitter and a receiver).
It is in this sense that Salecker
and Wigner~\cite{wign} consider the clock.
As mentioned, they also had in mind a rough picture
in which space-time could be in principle operatively
defined by a network of such free-falling clocks,
providing a material reference system~\cite{rovellimrs}.
If this (as it might well be)
was the proper way to obtain an operative definition of
space-time, one would obviously be led to consider each of the
clocks in the network to be extremely small and light.
In general a rather natural intuition is that the
ideal clocks to be used in the measurement of
a gravitational observable should be very light,
in order not to the disturb the observed quantity.
The same of course holds for all other devices
used in a gravitational measurement.
How light all these devices should be might depend on the
intended scale/sensitivity at which the measurement is
performed; for the operative definition of Planckian
distances one would expect that, since even tiny disturbances
would spoil the measurement, this ideal devices should be very
light, but the correct 
quantum gravity would be needed for a definite answer.

The criticism of the Salecker-Wigner limit expressed
in Ref.~\cite{anos}, was essentially based on two observations.
One of the observations, which I will address in the next two
sections, was based on the idea that
it might be possible to avoid the $\sqrt{T_{obs}}$ dependence
characteristic of the Salecker-Wigner limit.
The other observation, which I want to address in this section,
was based on the fact that
the data already available from 
the {\it Caltech 40-meter interferometer}
(the same here used in Section~2 to
set bounds on simple models of fuzziness)
imply that the effective clock mass to be used in the
Salecker-Wigner formula would have to be larger than 3 grams,
which the authors of Ref.~\cite{anos} felt to be to high a mass 
to be believable as a candidate mass of
fundamental clocks in Nature.
As underlined by the choice 
of observing~\cite{anos} that the 3-gram bound is comparable
to masses of wristwatch components,
this comment and criticism comes from taking literally
the Salecker-Wigner clock as a somewhat ordinary timing device.
This misses completely the point emphasized in the brief
review I have given above, {\it i.e.} that
the role of the effective Salecker-Wigner clock mass
cannot be taken literally as the mass of an ordinary
timing device: it is a more fundamental effective mass scale
characterizing the devices being used
(as clearly indicated by the fact that Salecker and Wigner
attribute to their conceptualization of
a ``clock'' the capability to transmit, receive
and process signals).
One must also consider that this idealized clock
was conceived as a device needed for a proper
operative definition of Planck-scale distances, and
therefore there is little to be gained from the intuition
of wristwatches and other ordinary timing devices.
The comment on the 3-gram bound given in Ref.~\cite{anos}
also fails to take into account the arguments,
which had already appeared in the
literature~\cite{bignapap,gacmpla,gacgrf98}
and have been here reviewed in Section~4,
concerning the need to interpret
the effective mass of the idealized Salecker-Wigner clock
as a fundamental but not necessarily
universal property of quantum gravity,
possibly depending on the type of length scales
involved/probed in
the experiment (as argued above for the
associated effective scale $max(M^*)$).
For experiments involving distance scales as large
as 40 meters, the result $max (M^*) > 3 grams$ seems perfectly
consistent\footnote{Perhaps a bound of the
type $max (M^*) > 3 grams$ would instead
be surprising if we had found it in experiments
defining Planckian distances in the spirit of the
type of networks of worldlines
considered by Salecker and Wigner (experiments which of course
are extremely far in the future if not impossible in principle).
Actually, it is quite daring to trust our feeling
of ``surprise'' when venturing so far from our present-day
intuition:~along the way to
the Planck scale we might be forced to change completely
our intuition about the natural world. For example,
on the subject of timing devices here of relevance
the interplay between gravitation and quantum mechanics
might even provide us new types of timing devices.
(One attempt to construct such new tools is discussed in
Ref.~\cite{dharam3QG} and some of the references therein.)}
with the idea that there should be some absolute bound
on $max(M^*)$ in any given quantum-gravity experimental
setup.
If experiments had given
a positive result (say, $max(M^*) \sim 2 grams$)
it would have not upset anything else we know abut the physical
world (only the most sensitive interferometers
would be sensitive to the effects of a Salecker-Wigner limit
with $max(M^*) \sim 2 grams$), but at the same time the fact that
it was instead found that $max (M^*) > 3 grams$
in experiments involving distance scales as large
as 40 meters should not surprise us
nor is it inconsistent with the arguments put forward by
Salecker and Wigner and followers.
Because of the present very early stage of development
of quantum gravity, we are at the same time
looking for the value (if any!)
of $max (M^*)$ and looking for an understanding
of what is the correct interpretation and the true physical
origin of such a bound on $max(M^*)$ in a quantum gravity
that would accommodate it at some fundamental level.

The points I discussed in this section
also clarify, within an explicit example,
the sense
in which the logic adopted in Ref.~\cite{anos} is inadequate
for the analysis of the conceptual framework set up by
Salecker and Wigner.
In Ref.~\cite{anos} the whole discussion of
the ``Salecker-Wigner clock'' remained 
strictly within the confines
of the intuition and the logic
of ordinary gravitation
and quantum mechanics, where we have nothing to learn.
The conceptual framework set up by
Salecker and Wigner instead treats the clock
in a way which, in as much
as it renounces to the idealization of a classical clock,
encodes one plausible 
departure from the ordinary laws of
quantum mechanics that could be induced by the process
of unification of gravitation with quantum mechanics.

\section{ON \space THE \space USE \space OF \space A \space
POTENTIAL \space WELL \space TO \space
REDUCE CLOCK-INDUCED UNCERTAINTY}

In Ref.~\cite{anos} the work of Salecker and Wigner
was also criticized by arguing that
it would be inappropriate to treat
the clock as freely moving, as effectively done in the derivation
of the Salecker-Wigner limit.
We were reminded in Ref.~\cite{anos} of the fact that
for a clock appropriately bound (say, by some ideal springs)
to another object in its vicinity
the uncertainty in the position of the clock
with respect to that object would
not increase with time, unlike the case of a free clock.

This observation completely misses the point of the
Salecker-Wigner limit. The uncertainty responsible for
the Salecker-Wigner limit comes from the uncertainty in
the relative position between the clock and the two bodies
whose distance is being measured (say, the distance between
the clock and the center of mass of the system composed of
the two bodies whose distance is being measured).
By binding in an harmonic potential the clock and an external
body one would not affect the nature of the Salecker-Wigner
analysis.
The position of the clock
(or, say, the center of mass of the system composed of
the clock and the external body)
relative to the two bodies
whose distance is being measured (or, say, relative to
the center of mass of the system composed of
the two bodies whose distance is being measured)
is still a free coordinate whose uncertainty contributes
directly to the uncertainty in our measurement of distance.
The uncertainty in this free coordinate
will spread according to the formula
\begin{eqnarray}
\delta x \geq \sqrt{ {\hbar T_{obs} \over 2}
\left( {1 \over  M_b}
+ {1 \over  M_c + M_{extra}} \right) }
~,
\label{deltawignANOS}
\end{eqnarray}
where $M_{extra}$ is the mass of the mentioned external body.
The $T_{obs}$ dependence necessary for
all the significant implications of the Salecker-Wigner analysis
is still with us.
Contrary to the claim made in Ref.~\cite{anos},
by binding in an harmonic potential the clock and an external
body, one does not truly eliminate the $T_{obs}$-dependent
uncertainty: one simply trades one source of $T_{obs}$-dependent
uncertainty for another essentially equivalent source.
This simply provides one more example of intuition for the $max (M^*)$
discussed in the preceding sections
(and in Refs.~\cite{bignapap,gacmpla,gacgrf98}), which in this context
would be identified with the inverse
of $min \{ 1/M_b + [1/(M_c + M_{extra})] \}$.
[In any case, as explained above, $M^*$ would plausibly not only
reflect the properties of the devices used for timing
but of the whole set of devices needed for the measurement of
distances.]

Whether or not there is a spring binding the clock and an external
body, as a result of the analysis of the Salecker-Wigner
measurement procedure we are still left with the intuition
that some fundamental (although perhaps dependent on the distance
scale which is to be measured~\cite{bignapap}) value for $max(M^*)$
might
be a prediction of quantum gravity and we are still left wondering
how large this $max(M^*)$ could be.
Perhaps when measuring large distances
with relatively low accuracy quantum gravity might allow us
to take rather large $M^*$ (which, if so desired,
one might effectively describe
in the language of Ref.~\cite{anos}
as the possibility to introduce
a rather heavy external body to be ``attached''
to the clock), but as shorter distances are probed
the disturbance of a large $M^*$ (or the introduction of heavy
bodies to which the clock would be attached) must eventually
become unacceptable. This is certainly plausible, but what could be
the value of $max (M^*)$ for measurements at a given distance scale?
The correct answer of course requires full quantum gravity
(because it must reflect the way in which the operative
definition of distances in codified in quantum gravity),
but we can try to gain some insight
by pushing further the experimental bounds on $max(M^*)$.
Even more complicated at the conceptual level
is the search of an analog of $M^*$
in attempts to operatively define a tight (perhaps Planck-length
tight) network of geodesic (world) lines,
in the spirit of ``material reference systems''~\cite{rovellimrs}
and of some of the comments
found in the work of Salecker and Wigner~\cite{wign}.
Is such a task to be required of quantum gravity?
How large/heavy could the clocks suitable for this task be?
Wouldn't it be paradoxical to consider the possibility of
attaching these free-falling clocks to some external bodies?
As already emphasized in Refs.~\cite{bignapap,gacmpla,gacgrf98}
there are several quite overwhelming open issues, but it
seems unlikely that we could gain some insight by
extrapolating {\it ad infinitum}
(as done in Ref.~\cite{anos})
from the intuition of
measurement-analysis ideas applicable to
rudimentary present-day experimental setups.

Before closing this section let me comment on another
scenario that some readers might be tempted to consider
as a modification of the potential-well proposal
put forward in Ref.~\cite{anos}. One might envisage
using some springs to connect the clock to one of the bodies
(say body $A$)
whose distance is being measured, rather than connecting
the clock to an external body.
This would assure that the uncertainty
in relative position between the clock
and that body $A$ does not increase with time, but it is easy to
verify that the disturbance
that this setup would introduce is of the same magnitude as the
uncertainty it eliminates.
In fact, the system composed of the clock and body $A$ would be
free. Essentially the uncertainty in the initial momentum
and position of the clock relative to the second body (body $B$)
would now be transferred to the body $A$ ``through the springs''.
This would introduce an uncertain disturbance to the distance
between body $A$ and body $B$ that is being measured, and the
disturbance is of course just of the same magnitude as the
uncertainty contribution arising in the original Salecker-Wigner
setup.
In addition, each time the (Salecker-Wigner-type) clock
emits a signal the corresponding uncertain recoil would be
transmitted through the spring to the body $A$.

\section{ON THE POSSIBILITY OF A FUNDAMENTALLY CLASSICAL CLOCK}

As an alternative possibility
to eliminate the $\sqrt{T_{obs}}$ dependence present
in the Salecker-Wigner limit, in Ref.~\cite{anos}
we are reminded of the fact that ordinary clocks
are immerged in a (thermal or otherwise) environment
that induces ``wave-function collapse''.
In fact, to extremely good approximation
these clocks behave classically.

Again this is a correct intuition derived from experience
with rudimentary (from a Planck-scale viewpoint) experimental
setups, which however (like the other points argued in 
Ref.~\cite{anos}) appears to be incorrectly applied to
the conceptual framework considered by Salecker and Wigner.
While ``environment-collapsed'' clocks (and other
environment-collapsed devices)
could be natural in ordinary contexts, it seems worth exploring
the idea that quantum gravity, as a truly fundamental theory
of space and time, would not resort (at an in-principle level)
to collapse-inducing environments for the operative definition
of distances. In any case, this is the expectation concerning
quantum gravity that is being explored through the
relevant Salecker-Wigner-motivated research line.
It also seems that quantum gravity, having to incorporate
an operative definition of distances applicable even in
the Planck regime,
would have some difficulties introducing at a fundamental
level the use of environments to collapse the wave function
of devices. How would such an environment look like for
the case in which one is operatively defining a
nearly-Planckian distance?
(and which type of environment
would be suitable for the operative definition
of a Planck-length-tight network of world lines?
how would such an environment be introduced in
the operative definition of a material reference system?)

Concerning the possibility of a fundamentally classical
clock in Ref.~\cite{anos}
the reader also finds what appears to be a genuinely
incorrect statement (not another example of ordinary intuition
inappropriately applied
to the forward-looking framework set up by Salecker and Wigner,
but simply a case of incorrect analysis).
In fact, Ref.~\cite{anos} appears to suggest
that the interactions among
the components of even a perfectly/ideally isolated clock
might induce classicality of the position of the
center of mass of the clock, which is the physical quantity
whose quantum properties lead to the Salecker-Wigner limit.
While the interactions among
the components should lead to the emergence of some classical
variables ({\it e.g.}, the variable that keeps track of time),
if the clock is ideally isolated interactions
among its components should not have any effect
on the quantum properties of
the position of the center of mass of the clock.
[This is certainly the case for some of the explicit
examples of ``toy clocks''
considered by Salecker and Wigner, one of which
is only composed of three free-falling particles!]

\section{CLOSING REMARKS}

From a conceptual viewpoint the analysis reported in Ref.~\cite{anos}
can be divided in two parts. In one part
a set of questions was raised and in the other part
tentative answers to these questions were given.
As this Letter emphasized, some of the questions
considered in Ref.~\cite{anos} are indeed the most fundamental
questions facing research based on the Salecker-Wigner limit.
However, all of these questions had already been raised in
previous literature
(see, {\it e.g.}, Refs.~\cite{bignapap,gacmpla,gacgrf98}).
These questions have been here compactly phrased as:
should quantum gravity predict a $max (M^*)$ and could this be
interpreted as the maximum acceptable mass of one or more devices?
how large could $max (M^*)$ be?
should $max (M^*)$ depend on the distance scales being probed?
should the idealization of a classical clock survive the transition
from ordinary quantum mechanics to quantum gravity?

While the questions considered are just the right ones,
the answers given in Ref.~\cite{anos} are incorrect.
In this note I have tried to clarify how those answers
are the result of inappropriately applying
the intuition of rudimentary (from a Planck-scale viewpoint)
measurement analysis
to the forward-looking framework set up by Salecker and Wigner.
The debate on the Salecker-Wigner limit must of course
continue until the above-mentioned
outstanding open questions get settled, but
(if the objective remains the one of getting ideas
on plausible quantum-gravity effects)
the only possibly fruitful way to
approach this problem
is the one of seeking the answers within the same
forward-looking framework where the questions arose.
Nothing more than what we already know
can be learned by assuming that the laws
of ordinary gravitation and quantum mechanics
remain unaltered all the way down to the Planck regime.
As emphasized here,
the logic of the line of research started by the work
of Salecker and Wigner is the one of applying
the language/structures we ordinarily use in those
physical contexts that we do understand
to contexts that instead would naturally
lie in the realm of quantum gravity,
and then exploring the consequences of removing one of
the elements of the ordinary conceptual structure of quantum
mechanics. The Salecker-Wigner study
(just like the Bohr-Rosenfeld analysis)
suggests that among these conceptual elements of
quantum mechanics the one that is most likely
to succumb to the unification of gravitation and quantum
mechanics is the requirement for devices to be treated
as classical. Removal of this requirement appears to guide
us toward some candidate properties of quantum gravity (not
of the ordinary laws of gravitation and quantum mechanics!),
which we can then hope to test directly in the laboratory
(as in some cases is actually possible~\cite{gacgwi,bignapap}).

Quite aside from the
subject of open issues in the study of the Salecker-Wigner
limit, I have also emphasized in this Letter that, contrary
to the impression one gets from reading Ref.~\cite{anos},
there is substantial motivation for
the phenomenological programme of interferometric
studies~\cite{gacgwi,bignapap} of distance fuzziness here
reviewed in Section~2, independently 
of the Salecker-Wigner limit (and independently
of the fact that, as clarified above,
the validity of this limit has not been seriously questioned).
As discussed in Section~2
(and discussed in greater detail
in Ref.~\cite{bignapap,polonpap}),
the general motivation for that phenomenological programme 
comes from a long tradition of ideas
(developing independently of the ideas related to
the Salecker-Wigner limit)
on foamy/fuzzy space-time,
and also comes from more 
recent work~\cite{aemn1,gacgrb,gampul,fordlightcone,adrian}
on the possibility that quantum-gravity
might induce a deformation of the dispersion relation that
characterizes the propagation of the massless particles
used as space-time probes in the operative definition
of distances.
It is actually quite important 
that this interferometry-based phenomenological programme,
as well as other recently-proposed quantum-gravity-motivated
phenomenological
programmes~\cite{polonpap,elmn,hpcpt,gacgrb,stringcogwi,grwlarge},
be pursued quite aggressively,
since the lack of experimental input
has been the most important obstacle~\cite{nodata}
in these many years of research on quantum gravity.

\vglue 0.6cm
\leftline{\Large {\bf Acknowledgements}}
\vglue 0.4cm
Part of this work was done while
the author was visiting the {\it Center for Gravitational
Physics and Geometry} of Penn State University.
I happily acknowledge
discussions on matters related to the subject of this Letter
with several members and visitors of the Center, particularly
with R.~Gambini and J.~Pullin.
I am also happy to thank
C.~Kiefer, for discussions on decoherence,
and D.~Ahluwalia, for feed-back on a first rough draft
of the manuscript.

\bigskip

\baselineskip 12pt plus .5pt minus .5pt

\end{document}